\documentclass[a4paper,fleqn,usenatbib]{mnras}

\usepackage[T1]{fontenc}
\usepackage{ae,aecompl}

\usepackage{units}
\usepackage{textcomp}
\usepackage{longtable}

\usepackage{graphicx}	
\usepackage{amsmath}	
\usepackage{amssymb}	
 \title[The circumstellar disk of FS~Tau~B]{The circumstellar disk of FS~Tau~B\\- A self-consistent model based on observations in the mid-infrared with \textit{NACO} -}

\author[Kirchschlager, Wolf and Madlener]{
Florian Kirchschlager,$^{1}$\thanks{E-mail: kirchschlager@astrophysik.uni-kiel.de}
Sebastian Wolf$^{1}$ and
David Madlener$^{2}$
\\
$^{1}$Kiel University, Institute of Theoretical Physics and Astrophysics,
 Leibnizstra\ss e 15, 24118 Kiel, Germany\\
$^{2}$Max Planck Institute for Radio Astronomy, Auf dem H\"ugel 69, 53121 Bonn, Germany
}

\date{Accepted XXX. Received YYY; in original form ZZZ}

\pubyear{2015}

\begin{document}
\label{firstpage}
\pagerange{\pageref{firstpage}--\pageref{lastpage}}
\maketitle

\begin{abstract}
Protoplanetary disks are a byproduct of the star formation process. In the dense mid-plane of these disks, planetesimals and planets are expected to form. The first step in planet formation is the growth of dust particles from submicrometer-sized grains to macroscopic mm-sized aggregates. The grain growth is accompanied by radial drift and vertical segregation of the particles within the disk. To understand this essential evolutionary step, spatially resolved multi-wavelength observations as well as photometric data are necessary which reflect the properties of both disk and dust.\\
We present the first spatially resolved image obtained with \textit{NACO} at the \textit{VLT} in the L$_\text{p}$ band of the near edge-on protoplanetary disk FS~Tau~B. Based on this new image, a previously published \textit{Hubble} image in H~band and the spectral energy distribution from optical to millimeter wavelengths, we derive constraints on the spatial dust distribution and the progress of grain growth. For this purpose we perform a disk modeling using the radiative transfer code MC3D. Radial drift and vertical sedimentation of the dust are not considered.\\
We find a best-fit model which features a disk extending from $\unit[2]{AU}$ to several hundreds AU with a moderately decreasing surface density and $M_\text{disk}=\unit[2.8\times10^{-2}]{M_\odot}$. The inclination amounts to $i=80^\circ$. Our findings indicate that substantial dust grain growth has taken place and that grains of a size equal to or larger than $\unit[1]{mm}$ are present in the disk. In conclusion, the parameters describing the vertical density distribution are better constrained than those describing the radial disk structure.
\end{abstract}




\section{Introduction}

The multiple system FS~Tau is located in the Taurus-Auriga star forming region at a distance of \mbox{$\unit[140]{pc}\pm\unit[20]{pc}$} (\citealt{Elias1978}). FS~Tau is a hierarchical triple-system, consisting of the narrow \mbox{T-Tauri}-binary FS~Tau~A (separation: $0.\hspace{-2.5px}\arcsec228 - 0.\hspace{-2.5px}\arcsec27$; \citealt{Simon1992}; \citealt{Hartigan2003}) and the young stellar object (YSO) FS~Tau~B at a projected separation of $\sim20\arcsec$ west. FS~Tau~A and FS~Tau~B are accompanied by a circumbinary and a circumstellar disk, respectively. The disk around FS~Tau~B is in the focus of this study. Alternative designations of FS~Tau~B are Haro~6-5B, HH~157, HBC~381, and IRAS~04189+2650.

The YSO has been classified as a Class I-II source (\citealt{Lada1987}) and has lost most of its original surrounding shell (\citealt{Yokogawa2001}). The disk is highly inclined \mbox{($i\approx67^\circ-80^\circ$;} \citealt{Krist1998}; \citealt{Yokogawa2001}) and obscures the central star at shorter wavelengths. Consequently, the disk appears as a bipolar nebula in the near-infrared (NIR), separated by an opaque band with a length of $3\arcsec-4\arcsec$ (\citealt{Padgett1999}) and a position angle of $144^\circ-150^\circ$. Since the disk is not orientated exactly edge-on and the dust particles potentially scatter non-isotropically, the two wings of the nebular structure differ in brightness. The disk mass has been constrained by several observations to $2\times10^{-3}\,\textrm{M}_\odot$ to $4\times10^{-2}\,\textrm{M}_\odot$ (\citealt{Dutrey1996}; \citealt{Yokogawa2001, Yokogawa2002}). Based on the observed low accretion rate, an age of $3.6\times10^{5}-2.4\times10^{6}$ years has been deduced (\citealt{Yokogawa2002}). Moreover, the object features a bipolar jet with a perpendicular orientation towards the opaque band (\citealt{Mundt1984}). 

Because of its low age and distance, FS~Tau~B is predestined to investigate the growth of dust grains in the context of planet formation. Due to its high inclination, the disk acts as a natural coronagraph reducing observational difficulties and artefacts common in coronagraphy. Furthermore, the vertical disk structure can be observed in the NIR without major disturbances by direct radiation from the stellar source. Comparable YSOs studied in the recent past are e.$\,$g.,~the Butterfly~star IRAS~04302+2247 (\citealt{Wolf2003Butterfly}; \citealt{Graefe2013}), CB~26 (\citealt{Sauter2009}), DG~Tau~B (\citealt{Kruger2011}), and HH~30 (\citealt{Madlener2012}).

Since the optical depth decreases in general with increasing wavelength, observations in the mid-infrared (MIR) allow the investigation of deeper regions of the disk and thus the study of thermal \mbox{re}emission of warm dust closer to the midplane. In this context, spatially resolved multi-wavelength observations are required to reduce ambiguities in the data analysis which exist due to the lack of knowledge regarding the dust density, chemical composition, and grain size distribution. In addition, observations in the MIR potentially provide constraints for dust particles in deeper layers and thus for possible settling of larger grains \mbox{(e.$\,$g.,~\citealt{Pinte2008};} \citealt{Graefe2013}).
 
While hot dust and scattered stellar light can be readily observed at NIR wavelengths, spatially resolved MIR observations tracing warm dust are rare. This applies also to the system of FS~Tau~B which was observed with the \textit{Hubble Space Telescope} (\textit{HST}) in the optical and NIR domain (\citealt{Krist1998}; \citealt{Padgett1999}) and with lower resolution at millimeter wavelengths (\citealt{Dutrey1996}; \citealt{Yokogawa2002}).

The aim of this study is to investigate the density distribution of the protoplanetary disk of FS~Tau~B and to constrain the evolutionary stage of grain growth. We present a new observation in the L$_{\text{p}}\,$band ($\lambda=\unit[3.74]{\text{\textmu} m}$) with $\sim0.\hspace{-2.5px} \arcsec1$ resolution (Sect. \ref{chap2}). Previously published observational data are summarised and presented in Section \ref{1234321}. The modeling campaign is based on the new observation in the MIR, a high resolution image obtained with \mbox{\textit{NICMOS/HST}} in the NIR, and published photometry data (Sect.~\ref{chap3}). The results are presented and discussed in Section~\ref{chap4}.

\section{Observation and data reduction}
 \label{chap2}
 \begin{figure} 
\hspace*{-1.0cm}\includegraphics[width=0.86\linewidth, angle=270]{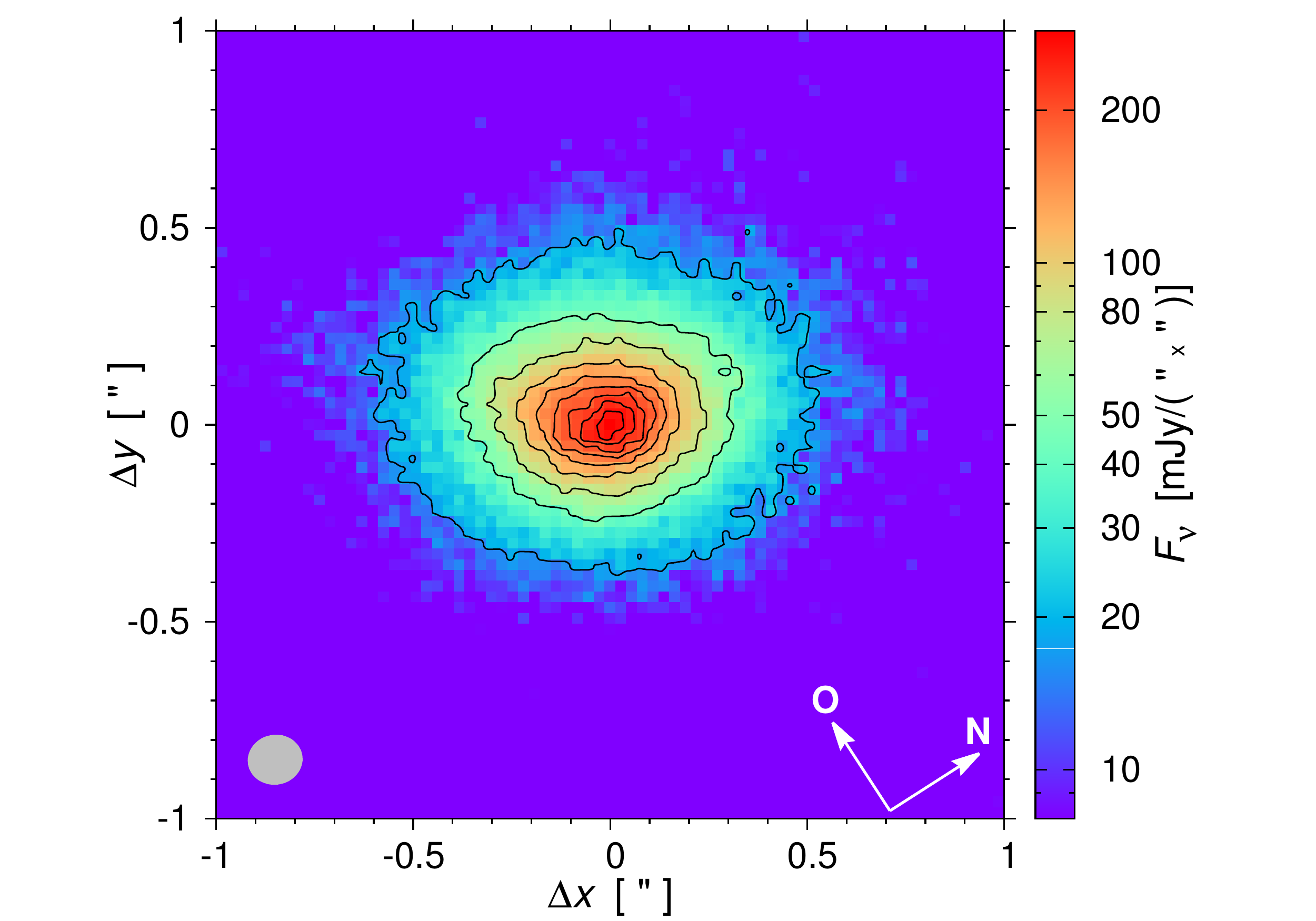}
\caption{Map of FS~Tau~B taken with \textit{NACO} in L$_\text{p}$~band ($\lambda=\unit[3.74]{\text{\textmu} m}$). The image was rotated by an angle of $-57^\circ$ in order to align the major axis of the elongated object with the horizontal image-axis. The contours correspond to flux densities in linear steps of $\unit[10]{\%}$, from $\unit[10]{\%}$ to $\unit[90]{\%}$ of the maximum flux density. The color scale is logarithmic. The ellipse in the bottom left corresponds to the FWHM of the PSF. The position angle of the major axis of the PSF is $PA=155^\circ\pm10^\circ$.}
\label{map_Image_LBand} 
 \end{figure}
FS~Tau~B was observed in service mode on three nights in December 2012 and January 2013 \mbox{[program} no.: 090.C-0207(A), PI: F. Kirchschlager] with the \textit{NACO} adaptive optics instrument, mounted on the UT4 at the \textit{Very Large Telescope} (\textit{VLT}, \citealt{Lenzen2003}; \citealt{Rousset2003}). The source was imaged with the L27-camera (pixel scale\,=$0.\hspace{-2.5px} \arcsec02712$ pixel$^{-1}$) in the L$_{\text{p}}\,$band \mbox{($\lambda=\unit[3.74]{\text{\textmu} m}$).} The functions \textit{Cube Mode} and \textit{Auto-Jitter} were used within a square field with a side length of $\Delta \phi=24\arcsec$. A single exposure lasted $t_\text{B}=\unit[0.25]{s}$. For each detector position $N_\text{B}=92$ single exposures were taken and the sum of detector positions for all three nights was $N_{\text{J}}=112$. These settings yield a total integration time of $t_\text{total}=\unit[2576]{s}\approx\unit[43]{min}$. The seeing during all observations was better than $1\arcsec$ and the relative airmass remained \mbox{below $1.6$.} The \mbox{L band standard star} FS~117 \mbox{(B216-b9)}, located at an angular distance of $0.5^\circ$ from FS~Tau~B, was used as reference. This point-like source has mag$_{\text{L}}=9.75^\text{mag}\pm0.01^\text{mag}$ (\citealt{Leggett2003}).
\begin{figure} 
 \includegraphics[height=1.0\linewidth, angle=270]{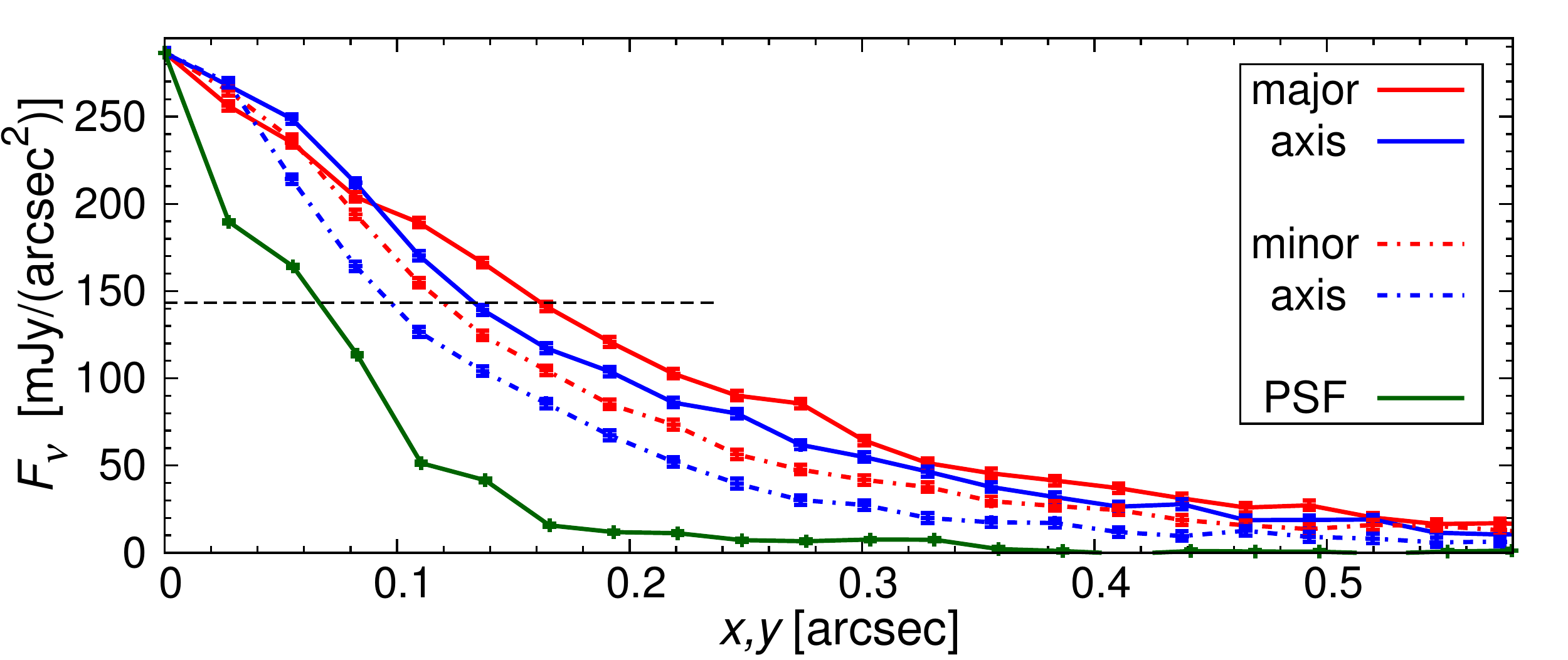}
\caption{Radial profiles of the observation of FS~Tau~B in L$_\text{p}$~band, centered on the pixel with maximum flux density. The flux densities as a function of distance $x$ ($y$) are determined as cuts along the major (solid lines) and minor axis (dashed lines). The radial profile along the major axis of the PSF is scaled for comparison. The dashed horizontal line marks the half maximum value to derive the FWHM.}
\label{img_rad_prof} 
 \end{figure}

The exposures were corrected for dark current and flat fielded. Bad pixels and cosmic ray artefacts were substituted by the median of the surrounding pixels, and all single exposures were added with respect to the jitter positions. The photometric calibration was performed by estimating the total count rates of the target and the reference star by using the average of all exposures. Different exposure times of the observations were taken into account. The total flux of FS~Tau~B in the L$_{\text{p}}\,$band amounts to $F_\nu=\unit[53.6]{m Jy}\pm\unit[1.1]{m Jy}$. This value matches well with the course of the spectral energy distribution (SED; see~Fig.~\ref{SED_bestes_Modell}). The uncertainty comprises the noise in the map of FS~Tau~B and FS~117 as well as the uncertainty of the flux of the reference star.

The reduced image was rotated by an angle of $-57^\circ$ in order to align the major axis of the object with the horizontal axis. The resulting map of FS~Tau~B in L$_\text{p}$~band ($\lambda=\unit[3.74]{\text{\textmu}m}$) is presented in Figure~\ref{map_Image_LBand} where the source appears as a single elongated object with a ratio of major to minor axis of $1.3$. The object is centered on the pixel with peak flux density. The position angle of the major axis is $PA=147^\circ\pm5^\circ$, a value that agrees well with literature data for the orientation of the disk plane (Tab.~\ref{Table_PWFSTauB}). Therefore, the major axis of the observed object is interpreted\break as the disk plane. The extension of the $\unit[10]{\%}$-contour line of \mbox{$\sim0.\hspace{-2.5px}\arcsec8\times1\arcsec$} is smaller than in the NIR (\citealt{Padgett1999}). 

Radial profiles were determined along the major and minor axis of the observed brightness distribution (Fig. \ref{img_rad_prof}). The FWHM for FS~Tau~B along the major axis is $x_\text{FWHM}=0.\hspace{-2.5px}\arcsec29$ and for the minor axis $x_\text{FWHM}=0.\hspace{-2.5px}\arcsec22$ while the FWHM for the radial profiles of the PSF are $x_\text{FWHM}=0.\hspace{-2.5px}\arcsec135$ and $0.\hspace{-2.5px}\arcsec12$, respectively. Therefore, FS~Tau~B is spatially resolved in this L$_\text{p}$~band observation. Left and right wing of the major/minor axis deviate by less than $0.\hspace{-2.5px} \arcsec03$, indicating only negligible asymmetry.
\begin{table} 
\caption{Position angle $PA$ of the disk plane of FS~Tau~B.} 
\label{Table_PWFSTauB} 
\begin{tabular}{c l c } 
\hline\hline 
Position angle $PA$&Ref.&Note\\\hline 
$144^\circ\pm\phantom{1}3^\circ$ &\cite{Krist1998} & (1) \\
$147^\circ$\,\phantom{ $\pm17^\circ$}&\cite{Padgett1999} & (1) \\
$138^\circ\pm\phantom{1}7^\circ$ &\cite{Yokogawa2001} & (1) \\
$147^\circ\pm\phantom{1}5^\circ$ & this work & (1) \\
$150^\circ\pm10^\circ$ &\cite{Yokogawa2002} & (2) \\
$147^\circ\pm\phantom{1}8^\circ$ &\cite{Gledhill1986} & (3) \\\hline
\end{tabular}
\newline\textbf{Note:} (1) $PA$ of the dust disk; (2) $PA$ of the gas disk; (3) $PA$ corresponds to polarization angle in the optical wavelength range.
\end{table}

\section{Further observations}
\label{1234321}
For the modeling of the disk of FS~Tau~B, further observational data are required which are presented in this section.
\subsection{Spectral energy distribution}
\label{nr100} 
\begin{figure} 
\includegraphics[height=1.0\linewidth, angle=270]{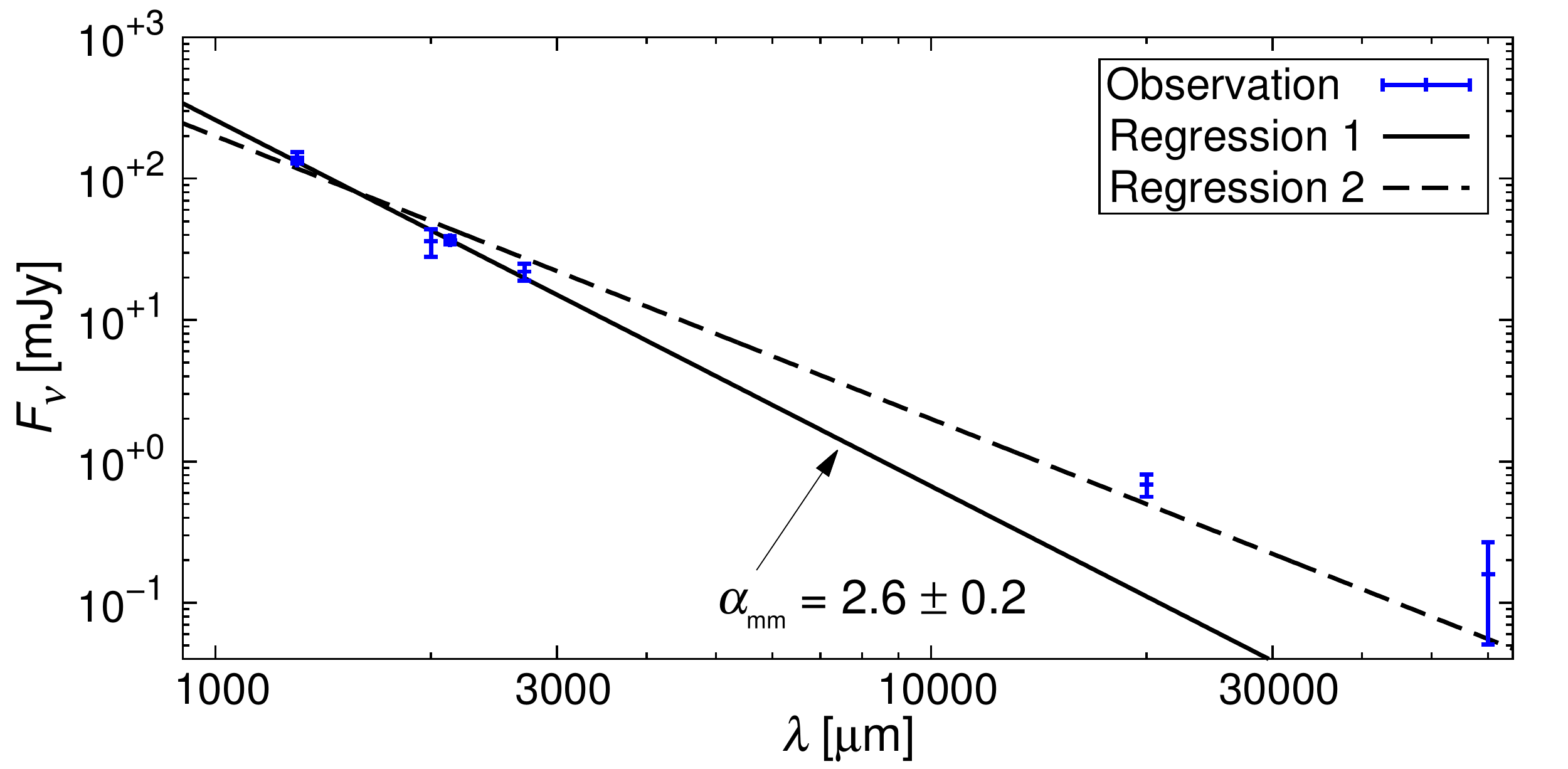}
\caption{Spectral index for the disk of FS~Tau~B. The lines are determined using the method of least squares. If the fluxes in the centimeter range are neglected, the slope is $\alpha_\text{mm}=2.6\pm0.2$ (solid line), otherwise $\alpha'_\text{mm}=2.0\pm0.2$ (dashed line).}
\label{FSTauB_Spektralindex} 
 \end{figure}

The photometric data points are summarized in Table~\ref{Table_SEDFSTauB} in the appendix and plotted in Figure~\ref{SED_bestes_Modell}. The dataset from \textit{Spitzer} (\textit{IRS/SST}) contains 366 data points in the wavelength range from $\unit[5.2]{\text{\textmu} m}$ to $\unit[37.9]{\text{\textmu} m}$. The known SED of FS~Tau~B comprises continuum fluxes from the optical up to the millimeter and radio range and shows some features:
\begin{itemize}
 \item A disk with high inclination as seen on the NIR~image (\citealt{Padgett1999}; Fig.~\ref{Padgett}) should cause higher fluxes in the scattering range of the SED compared to the maximum flux at wavelengths in the far-infrared. This is not the case for FS~Tau~B.\\[-0.3cm]
 \item Thermal \mbox{re}emission at MIR wavelengths suggests the presence of warm dust and therefore a small inner disk radius.\\[-0.3cm]
\item The \textit{IRS/SST} dataset shows an absorption feature which indicates the presence of crystalline silicate in the dust. Compared to other highly inclined disks (e.$\,$g.,~CB~26, HH~30, Butterfly~star), the feature is less pronounced for FS~Tau~B.\\[-0.3cm]
\item The regression line for the \mbox{(sub-)}millimeter range of the SED is presented in Figure \ref{FSTauB_Spektralindex}. The slope has a value of $\alpha_\text{mm}=2.6\pm0.2$ and is significantly smaller than in the case of dust in the interstellar medium (ISM; $\alpha_\text{mm}\sim3.7$; \citealt{WeingartnerDraine2001}) which suggests that grain growth has taken place in the disk (e.$\,$g.,~\citealt{Natta2007}). Moreover, when the regression also includes observed fluxes at $\lambda=\unit[2]{cm}$ and $\unit[6]{cm}$, the slope reduces further to $\alpha'_\text{mm}=2.0\pm0.2$. However, these observations might be contaminated by free-free-radiation of the jet (e.$\,$g.,~\citealt{Marti1993}; \citealt{Pety2006}). 
\end{itemize}
 \begin{figure} 
\includegraphics[height=1.0\linewidth, angle=270]{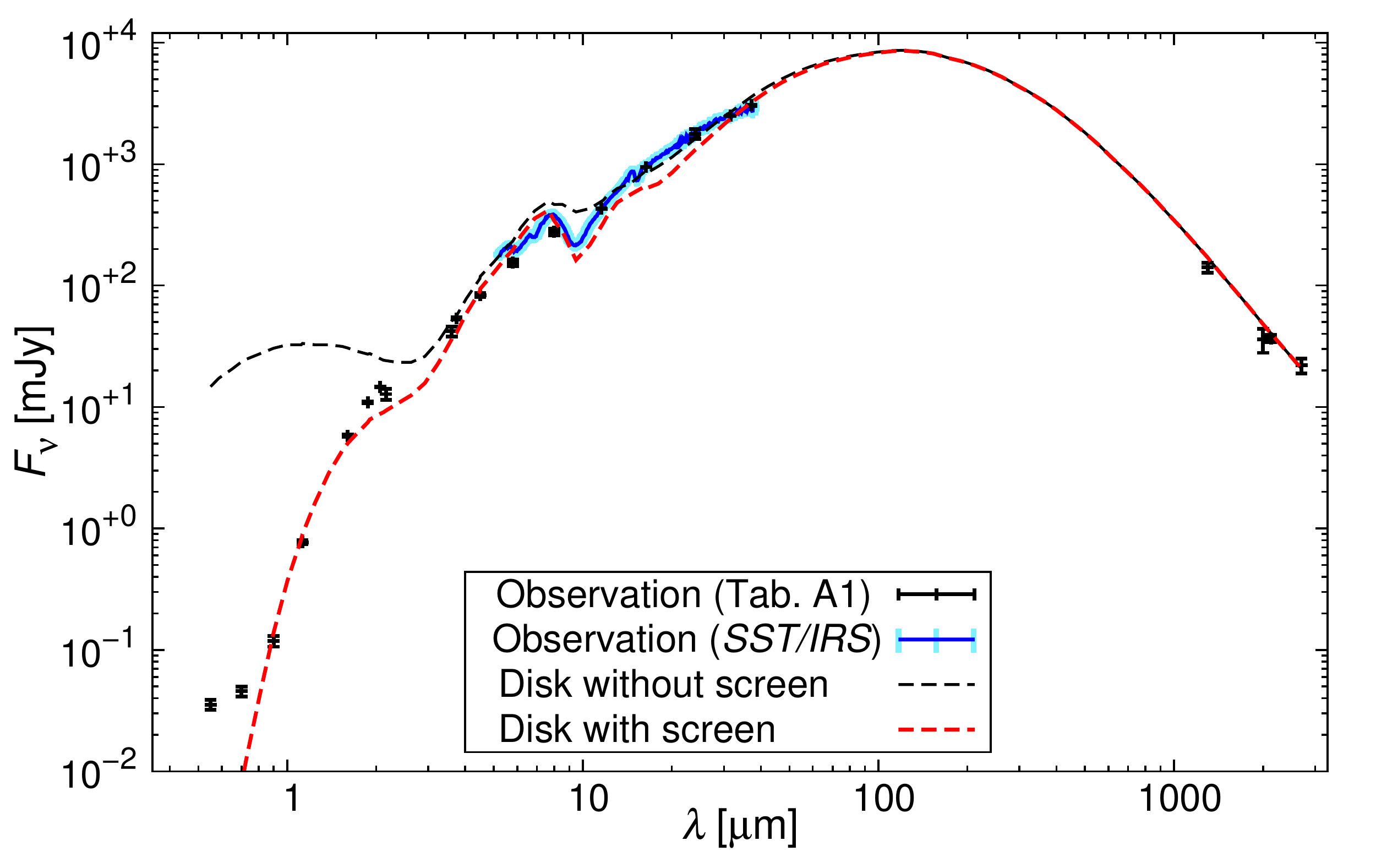}
\caption{Photometric data of FS~Tau~B and the SED of the best-fit model found in Section~\ref{chap4} (red dashed line). The black dashed line shows the SED of the same model without extinction screen ($A_{\text{V}}=0$).}
\label{SED_bestes_Modell} 
 \end{figure}

\subsection{Observation in the near-infrared}
\begin{figure} 
\hspace*{-1.0cm}\includegraphics[width=0.86\linewidth, angle=270]{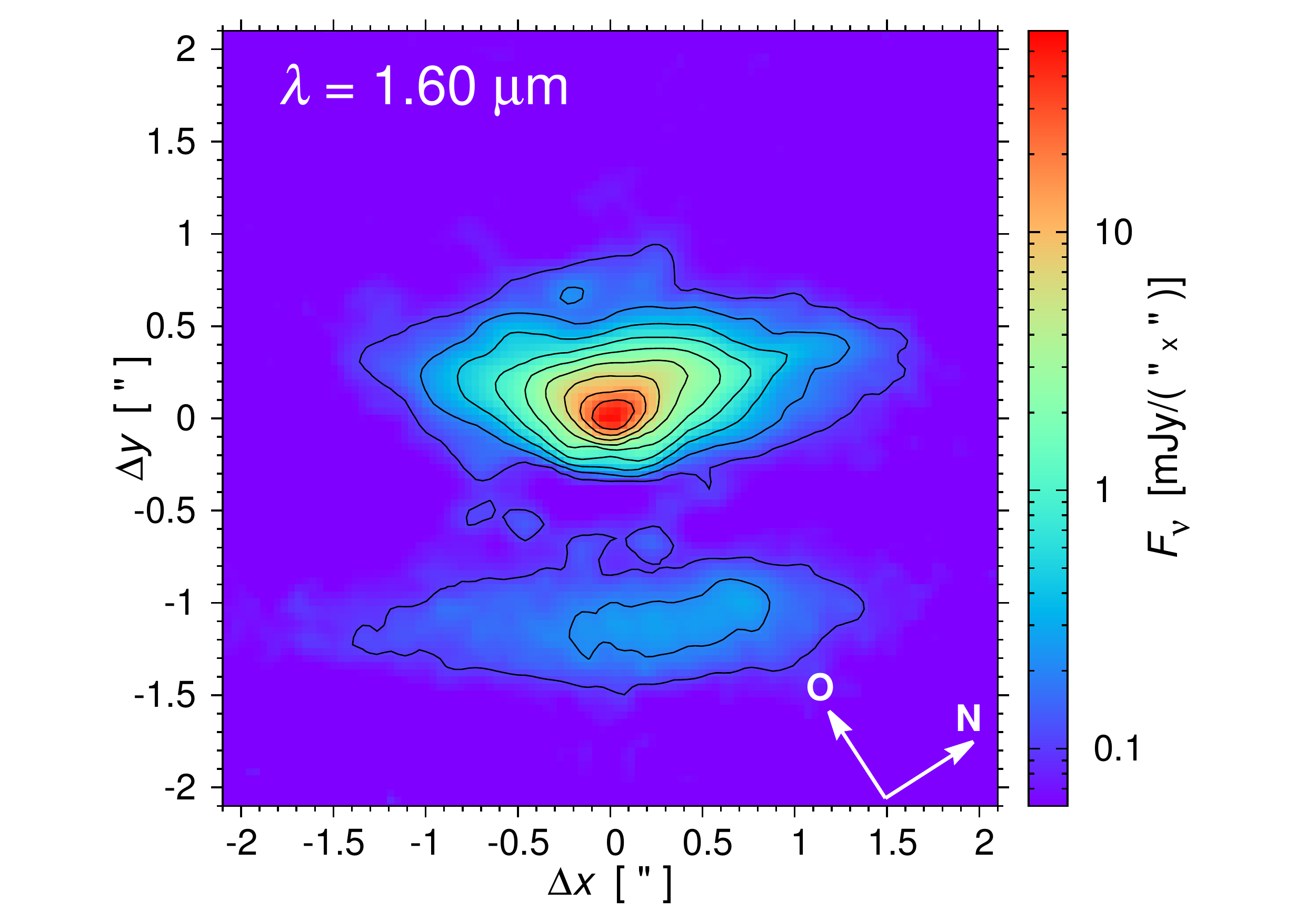}
\caption{Map of FS~Tau~B taken with the \textit{HST} in the F160W-filter ($\lambda=\unit[1.60]{\text{\textmu} m}$). The contours correspond to flux densities in linear steps of $\unit[10]{\%}$, from $\unit[10]{\%}$ to $\unit[90]{\%}$ of the maximum flux density. The color scale is logarithmic (adopted from \citealt{Padgett1999}; private communication with W. Brandner).}
\label{Padgett} 
 \end{figure}

FS~Tau~B was observed in the NIR with the NIC2 camera of the instrument \textit{NICMOS/HST} on October 29th, 1997 (\citealt{Padgett1999}) in the filters F110W, F160W, F187W, and F205W (instrument description: \citealt{Thompson1998}). The detailed data reduction procedure is described in \cite{Padgett1999}. The object is seen in the maps as a bipolar nebula. The bipolar appearance decreases with increasing wavelength which is consistent with our observation in L$_\text{p}$~band where only a single lobe can be seen (Fig.$\,$\ref{map_Image_LBand}). In Figure~\ref{Padgett} the image at $\lambda=\unit[1.60]{\text{\textmu} m}$ taken from \cite{Padgett1999} is presented which was used for modeling in the following section.

\section{Disk modeling}
 \label{chap3}
In this section our approach to reproduce the appearance of the protoplanetary disk of FS~Tau~B is discussed. In the following, the applied software, the disk and dust model, as well as the modeling procedure are introduced.

\subsection{Radiative transfer}
We used the program \texttt{MC3D} for our radiative transfer simulations which is based on the Monte Carlo method (\citealt{Wolf1999}; \citealt{Wolf2003}). The code implements the temperature-correction technique described by \cite{BjorkmanWood2001}, the absorption concept of \cite{Lucy1999}, and the scattering scheme by \cite{CashwellEverett1959}. The radiation field is described by the Stokes formalism ($I$, $Q$, $U$, $V$; \citealt{Stokes1852}; \citealt{Chandrasekhar1946}). The software first simulates the temperature distribution and then uses this information to calculate the SED, as well as spatially resolved scattering and \mbox{re}emission maps for the given dust density distribution.

\subsection{Model for FS~Tau~B}
\label{2015Jan19}
Our chosen model consists of three components: The central star, a flared circumstellar disk and an extinction screen between the system and the observer. Observations in the millimeter domain by \cite{Yokogawa2001} and synthetic imaging in the NIR by \cite{Stark2006} showed that FS~Tau~B already lost the shell which enveloped the disk during the earliest stages. Therefore, a shell is not considered in our modeling approach of FS~Tau~B.

The disk is heated by a source which is assumed to be a black body, characterized by its effective temperature $T_\text{heat}$ and radius $R_\text{heat}$, resulting in a corresponding luminosity $L_\text{heat}$. Both quantities ($T_\text{heat}$, $R_\text{heat}$) were varied during the modeling process. The heating source is composed of the central star and further contributions such as viscous friction and accretion shocks.

For the density distribution we adopt the canonical flared disk parametrization that depends on the radial distance from the star and the distance from the disk midplane (\citealt{ShakuraSunyaev1973}),
\begin{equation}
 \rho(R,z)=\rho_0 \left(\frac{R}{R_{100}}\right)^{-\alpha} \exp{\left(-\frac{1}{2}\left[\frac{z}{h(R)}\right]^2\right)} .\label{Shakura} \end{equation}
Here, $R$ is the cylindrical distance from the center and $z$ the distance from the disk midplane. The factor $\rho_0$ fixes the total disk mass given the radial boundaries. The scale height 
\begin{equation}
h(R)=h_{100}\left(\frac{R}{R_{100}}\right)^{\beta}
\end{equation} 
encodes the flaring of the disk. The inner and outer radius $R_\text{in}$ and $R_\text{out}$, the geometrical parameters $\alpha$ and $\beta$, the scale height $h_{100}$ at the reference radius $R_{100}=\unit[100]{AU}$ as well as the dust mass $M_\text{dust}$ are in total six free parameters.

\begin{figure} 
\includegraphics[width=1.0\linewidth]{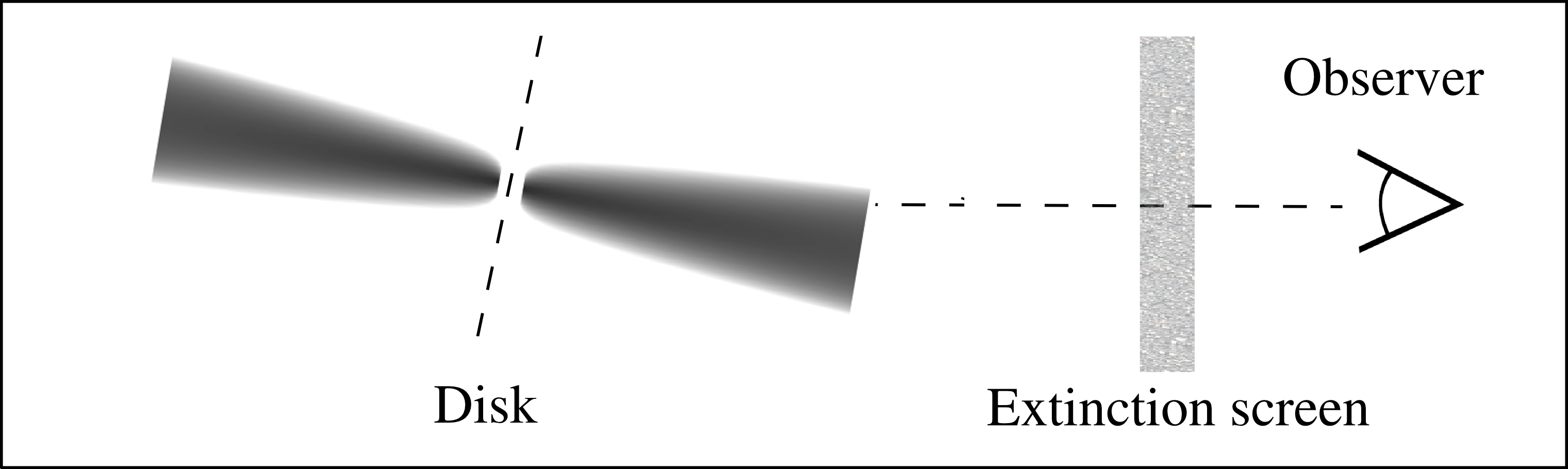}
\caption{Position of the extinction screen between disk and observer.\vspace*{-0.3cm}}
\label{Skizze_System} 
\end{figure}

FS~Tau~B is located in the Taurus-Auriga star forming region and surrounded by a complex accumulation of material. To take the ISM into account as a source of extinction, we adopt an extinction screen between disk and observer which attenuates the radiation of the object as a function of wavelength, see Figure~\ref{Skizze_System}. A further motivation lies in our finding that the SED cannot be reproduced by a highly inclined disk $(i>60^\circ)$ without an extinction screen ($A_{V}=0$). The extinction properties of interstellar dust grains are assumed for this screen and its thickness is characterized by the optical extinction $A_{V}\approx1.086\,\tau_{V}$ where $\tau_{V}$ denotes the optical depth in V~band ($\lambda=\unit[0.548]{\text{\textmu} m}$). The observables calculated by radiative transfer are modified by the wavelength dependent absorption of the screen. In Section \ref{chap4} we show that thermal \mbox{re}emission from the extinction screen is negligible.

The distance towards FS~Tau~B corresponds to the distance of the Taurus-Auriga star-forming region at $d=\unit[140]{pc}\pm\unit[20]{pc}$.
The disk inclination $i$ is treated as another free parameter.
\subsection{Dust model}
The dust grains in our disk model are assumed to be spherical and composed of $\unit[62.5]{\%}$ astronomical silicate and $\unit[37.5]{\%}$ crystalline graphite with a total bulk density of $\rho=\unit[2.7]{g\,cm^{-3}}$ (\citealt{WeingartnerDraine2001}). The dust grains follow the size distribution \mbox{$\mathrm{d}n\propto a^{-q}\,\mathrm{d}a$} with exponent $q=3.5$ (\citealt{Mathis1977}) with grain radii varying in the interval \mbox{$a\in \left[\unit[5]{nm}, a_\text{max}\right]$}. Since the spectral index already indicates the presence of larger particles than found in the ISM, six discrete values for the maximum grain radius are considered: $a_\text{max}=\unit[0.25]{\text{\textmu} m}, \unit[1]{\text{\textmu} m}, \unit[5]{\text{\textmu} m}, \unit[20]{\text{\textmu} m}, \unit[100]{\text{\textmu} m}$, and $\unit[1]{mm}$. A spatial variation of grain size properties driven by segregation processes like dust settling is not considered. The optical properties of the particles are calculated with the program \texttt{miex} which is based on the theory of Mie-scattering (\citealt{Mie}; \citealt{WolfVoshchinnikov04}). To reduce computational time and memory requirements for radiative transfer simulations, we applied the approximation by \cite{Wolf2003b} which replaces the optical properties of a single grain with radius $a$ by the weighted mean of optical properties for the entire grain population.
\subsection{Quality of the fit}
\label{Nr683}
To search for the best-fit model we have to compare the simulated quantities to the observational data. Therefore, we use the concept of $\chi^2$ minimization with
\begin{align}
 \chi^2\propto\left(\chi^2_\text{SED}+\sum_{k=1}^{N}g_k\,\chi_{k}^2\right),\label{2124}
\end{align}
where $\chi^2_\text{SED}$ and $\chi_{k}^2$ are the individual contributions of the SED and the $k$-th of $N$ maps, weighted with $g_k$, respectively.
Here, $N=2$ as we consider two maps at $\lambda=\unit[1.60]{\text{\textmu} m}$ and $\unit[3.74]{\text{\textmu} m}$. For both, only the radial profiles along the major and minor axes are taken into account. Each flux of the SED and brightness profiles are normalized by dividing the difference between observed and modeled flux with their individual uncertainties. The maps are equally weighted ($g_1=g_2$) and the $g_k$ are chosen in a way that $\chi^2_\text{SED}$ and the sum of the contributions of the two maps get in balance.

\subsection{Modeling strategy}
The parameter space of the described model is \mbox{11-dimen}-sional. All parameters and their ranges are presented with the values for the best-fit model in Table~\ref{TabParameterFSTauB}. The chosen intervals are based on restrictions by observational data and modeling of similar objects (e.$\,$g.,~\citealt{Sauter2009}; \citealt{Graefe2013}).

First the temperature distribution is calculated for each parameter set, then the SED and the scattering- and \mbox{re}emission maps for $\lambda=\unit[1.60]{\text{\textmu} m}$ and $\unit[3.74]{\text{\textmu} m}$ are generated. The maps are convolved with an elliptical Gaussian function and the radial profiles are extracted along the major and minor axis. The simulated quantities are then compared to the observations and $\chi^2$ is calculated according to Equation~\ref{2124}.

The parameters $i$ and $A_{V}$ are not affected by the results of the radiation transport simulations and can thus be altered subsequently. The remaining parameters span a \mbox{9-dimensional} parameter space. In order to find a model that reproduces most of the observational data, the method of \cite{Sauter2009} is applied and the parameters are fitted iteratively. At first the range of an individual parameter is sampled in four coarse steps. Then the two best values are selected and the procedure is continued to the next parameter. This sequence is repeated several times using random parameter order. The stepping of each parameter is refined in a smaller range in each sequence, based on the results of the previous sequence. The procedure ends when a sequence delivers no better model.

The uncertainties of each parameter are calculated by determining the range this parameter can be altered without exceeding the $\chi^2$ of the best-fit model by more than $\unit[10]{\%}$. There is no mathematical reason supporting this value, but its applicability has been proven within our study because deviations larger than $\unit[10]{\%}$ generally give inferior results. The confidence intervals determined with this approach are unsymmetrical unlike the usual $1\sigma$-interval.

\section{Results}
\label{chap4}
The resulting best-fit model reproduces the key characteristics of the observational data (Tab.$\,$\ref{TabParameterFSTauB}).
The following sections discuss the parameters of the best-fit model (Sect.$\,$\ref{Describe}), the simulated SED (Sect.$\,$\ref{AbschnSED12}, Fig.$\,$\ref{SED_bestes_Modell}), brightness maps (Sect.$\,$\ref{AbschnittKarten}, Fig.$\,$\ref{FSTAUB_results_maps}) and the temperature distribution in the disk midplane (Sect.$\,$\ref{Temp2000}).
 
\subsection{Disk parameters}
\label{Describe}
 \begin{table} 
 \caption{Parameter space and values of the best-fit model.}
 \label{TabParameterFSTauB} 
\centering
 \begin{tabular}{c l c c} 
\hline\hline 
 \multicolumn{2}{c}{Parameter}&Parameter space&Best-fit model\\\hline 
&&&\\[-0.25cm]
 $R_\text{in}$ 		&$[$AU$]$	&$0.1-10$\phantom{.}& $\phantom{000}2.0_{-0.3}^{+\,1.2}$ 	 \\
&&&\\[-0.25cm]
 $R_\text{out}$ 		&$[$AU$]$	&$150-400$ & $\phantom{0.0}350_{-150}^{+\,\phantom{0}50}$ \\
&&&\\[-0.25cm]
 $ h_{100}$ & $[$AU$]$ &\phantom{2}$5-25$& $\phantom{.000}10_{-1}^{+\,2}$\\
&&&\\[-0.25cm]
 $\alpha$ & &$1.0-4.0$& $\phantom{000}2.1_{-0.6}^{+\,0.5}$ \\
&&&\\[-0.25cm]
 $\beta$&&$1.0-2.0$& $\phantom{00}1.20_{-0.01}^{+\,0.06}$ \\
&&&\\[-0.25cm]
 $M_\text{dust}$ 	&$[\textrm{M}_{\odot}]$	&$10^{-6}-10^{-2}$& $\phantom{000}\left(2.8_{-0.7}^{+\,0.3}\right)\times10^{-4}$\\
&&&\\[-0.3cm]
 $a_\text{max}$ 	&$[\text{\textmu}\textrm{m}]$&\phantom{1}$0.25-1000$\phantom{.}& $1000\phantom{0.}$\\\hline
&&&\\[-0.25cm]
 $i$ &$[{}^\circ]$&$60-90$ & $\phantom{0.}80_{-2}^{+\,1}$ 	 \\
&&&\\[-0.25cm]
 $A_{\text{V}}$&&\phantom{0}$0-20$& $12\phantom{.}$ \\ \hline
&&&\\[-0.25cm]
 $ T_\text{heat}$&$[$K$]$&$3000-8000$ & $\phantom{0.}7000_{-600}^{+\,500}$ 	 \\
&&&\\[-0.25cm]
 $ R_\text{heat}$&$[\textrm{R}_{\odot}]$	&$0.8-3.5$& $\phantom{000}2.1_{-0.5}^{+\,0.4}$ 	 \\\hline 
 \end{tabular}
 \end{table} 
In this section, the disk parameters of the best-fit model are discussed separately.

The inner edge of the disk, as seen in thermal MIR emission maps, amounts to $R_\text{in}=\unit[2.0_{-0.3}^{+\,1.2}]{AU}$, indicating the disk has no extended inner hole as observed for several other objects (e.$\,$g.,\citealt{Andrews2011}; \citealt{Graefe2011}). For the given central object, the sublimation radius is $R_\text{sub}\approx\unit[0.35]{AU}$ (\citealt{Whitney2004}). The high MIR~fluxes imply the presence of warm dust located relatively close to the star.

Due to the low signal to noise ratio in the outer regions, the outer radius $R_\text{out}=\unit[350_{-150}^{+\,\phantom{0}50}]{AU}$ is only weakly constrained. The studies of \cite{Krist1998}, \cite{Padgett1999}, \cite{Yokogawa2001}, and \cite{Stark2006} obtained comparable values between $\unit[240]{AU}$ and $\unit[309]{AU}$ for the outer radius.

The scale height at $R_{100}=\unit[100]{AU}$, \mbox{$h_{100}=\unit[10_{-1}^{+\,2}]{AU}$}, the radial exponent $\alpha=2.1_{-0.6}^{+\,0.5}$, and the flaring exponent $\beta= 1.20_{-0.01}^{+\,0.06}$ are typical for a protoplanetary disk (e.$\,$g.,~IM~Lupi, \citealt{Pinte2008}; CB~26, \citealt{Sauter2009}). In addition, the criterion $\alpha=3(\beta-\frac{1}{2})$ from viscous accretion theory is fulfilled (\citealt{ShakuraSunyaev1973}). The exponent of the surface density distribution, $p=\alpha-\beta=0.9_{-0.6}^{+\,0.5}$, is consistent with theoretical (e.$\,$g.,~\citealt{Bell1997}) and empirical studies (e.$\,$g.,~\citealt{Kitamura2002}; \citealt{Andrews2007}) which find $0.5\lesssim p\lesssim 1$, indicating an monotonically decreasing density towards the outer edge. The surface density amounts to $\Sigma=\unit[16.3]{g\,cm^{-2}}$ and $\Sigma=\unit[1.1]{g\,cm^{-2}}$ at the distances $R=\unit[5]{AU}$ and $\unit[100]{AU}$, respectively.

The total dust mass in the disk is determined to $M_\text{dust}=\unit[2.8_{-0.7}^{+\,0.3}\times10^{-4}]{M_\odot}$, assuming compact, spherical, and homogenous dust particles. If the canonical value of $M_\text{gas}/M_\text{dust}=100$ is adopted (e.$\,$g.,~\citealt{Hildebrand1983}), we derive a total mass of $\unit[\sim2.8\times10^{-2}]{M_\odot}$. This value is in agreement with \cite{Dutrey1996}, \cite{Krist1998}, \cite{Yokogawa2002}, and \cite{Stark2006}, who obtained values for the total disk mass ranging from $M_\text{dust}=\unit[1\times10^{-2}]{M_\odot}$ to $\unit[4\times10^{-2}]{M_\odot}$. Moreover, the criterion found by \cite{Toomre1964} for a rotating disk indicates that the disk is gravitationally stable at all disk radii ($Q_\text{T}\gg1$).

The maximum grain radius of the best-fit model is $a_\text{max}=\unit[1]{mm}$, more than three orders of magnitude larger than the maximum grain radius of the ISM (\citealt{Mathis1977}). We therefore conclude that grain growth has taken place in the disk of FS~Tau~B. Large particles radiate effectively in the \mbox{(sub-)}millimeter wavelength range and shape the slope in this range (Fig.$\,$\ref{FSTauB_Spektralindex}). Since the next lower value in the parameter study is significantly smaller ($\unit[100]{\text{\textmu} m}$), $a_\text{max}$ is fixed to ${\unit[1.0]{mm}}$ and no uncertainties for this parameter are given. Although the derived value for the maximum grain radius is at the edge of the considered parameter space, the presence of larger dust particles cannot be deduced, as their effect on the observed SED is minor.

The inclination $i$ is constrained by the SED and the radial brightness profiles to $i=80^\circ {}_{-2^\circ}^{+\,1^\circ}$. This value is consistent with the edge-on orientation deduced from the NIR~maps. \cite{Krist1998}, \cite{Padgett1999}, \cite{Yokogawa2001}, and \cite{Stark2006} found inclinations ranging from $i=67^\circ$ up to $80^\circ$.

The central star has a spectral type of K$5\pm2$ (\citealt{White2004}) which corresponds to a stellar temperature of $T_\star=\unit[4400_{-400}^{+\,400}]{K}$. Using
evolutionary tracks of pre-main-sequence stars (\citealt{Siess2000}), a stellar mass of $M_\star=\unit[1.0]{M_\odot}$ to $\unit[1.4]{M_\odot}$ can be derived. With a temperature $T_\text{heat}=\unit[7000_{-600}^{+\,500}]{K}$ and radius $R_\text{heat}=\unit[2.1_{-0.5}^{+\,0.4}]{AU}$, we deduce a luminosity $L_\text{heat}=\unit[9.5_{-3.5}^{+\,2.9}]{L_\odot}$ for the heating source. If we assume that the derived best-fit luminosity $L_\text{heat}$ is composed of contributions from the central star ($L_\star$) and accretion ($L_\text{acc}$) only, we can derive the mass accretion rate using the relationship $L_\text{acc}=\frac{G\,M_\star}{2\,R_\star}\frac{\text{d}M}{\text{d}t}$, where $G$ is the gravitational constant. The mass accretion rate is then constrained to the interval $\unit[3.2\times10^{-7}]{\nicefrac{\text{M}_\odot}{yr}}-\unit[1.2\times10^{-6}]{\nicefrac{\text{M}_\odot}{yr}}$ which is comparable to previous studies ($\unit[9.3\times10^{-8}]{\nicefrac{\text{M}_\odot}{yr}}-\unit[6.2\times10^{-7}]{\nicefrac{\text{M}_\odot}{yr}}\,$; \citealt{Yokogawa2002}).

The optical extinction $A_\text{V}$ is very sensitive to other parameters such as the inclination, disk mass or inner radius. 
Since the uncertainty of this parameter is dominated by the variation of other parameters, the errors are not specified. In a fitting approach considering only the SED, models without extinction screen ($A_\text{V}=0$) can reproduce the SED very well. However, the inclination in these parameter sets is only $40^\circ-50^\circ$, thus contradicting information from observed images. On the other hand, a highly inclined disk yields a SED with larger fluxes than observed at wavelengths up to several micrometers. Following \cite{Sauter2009}, an extinction screen with $A_\text{V}=12$ is introduced which causes a wavelength-dependent reduction of these fluxes. The screen is optically thin at (sub-)millimeter wavelengths ($\tau_{\unit[1.3]{mm}}\approx8\times10^{-5}$). If we assume a dust temperature of $\sim\unit[20]{K}$ in the screen, the resulting flux amounts to $F_\nu(\lambda=\unit[1.3]{mm})\approx\unit[2]{mJy}$, two magnitudes lower than the observed flux. Thus, the thermal \mbox{re}emission of the extinction screen can be neglected.

In summary it can be concluded that parameters which describe the vertical density distribution ($\beta$, $h_{100}$) of the disk are better restricted than those influencing the radial disk structure ($R_\text{in}$, $R_\text{out}$, $\alpha$). 

 \begin{figure} 
 \includegraphics[height=1.0\linewidth, angle=270]{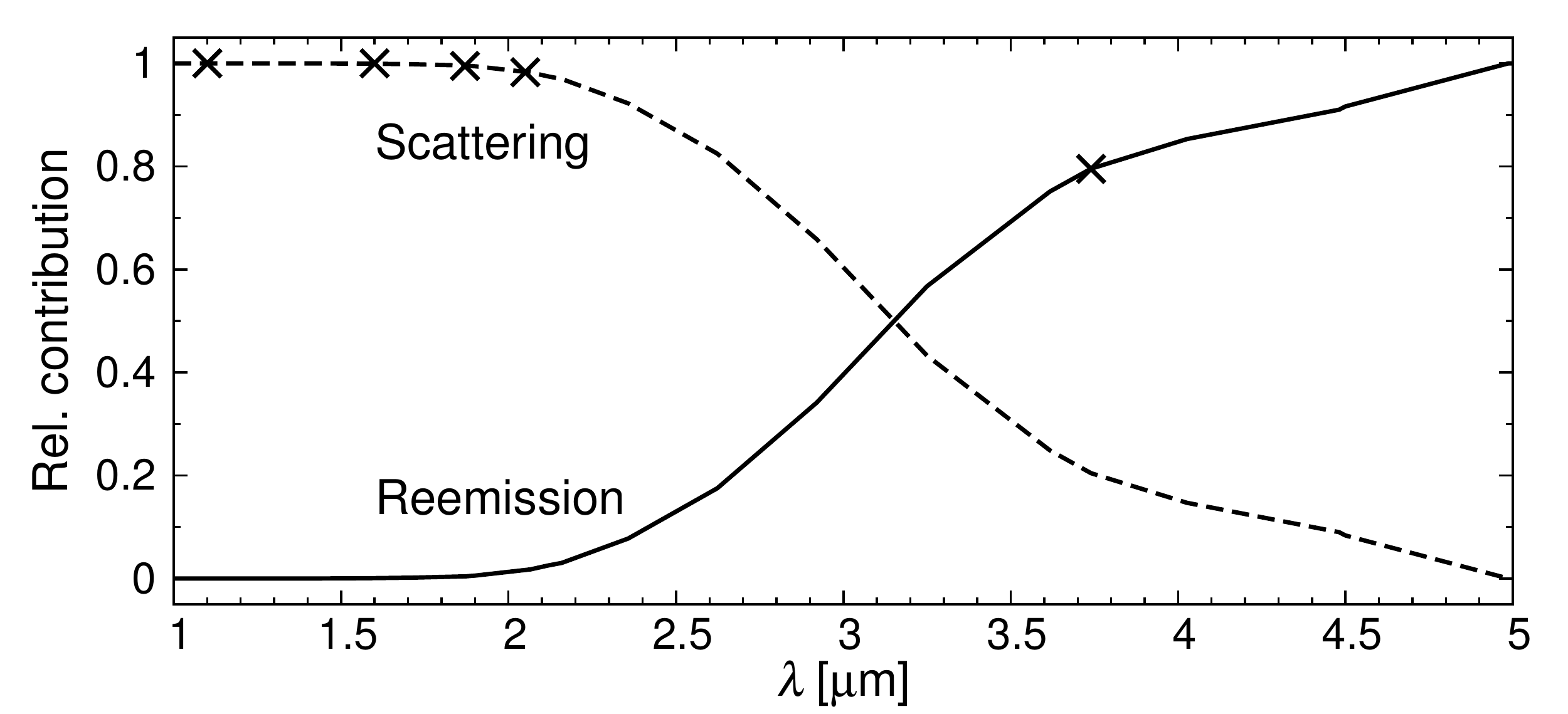}
\caption{Relative contribution of scattered and reemitted radiation to the total intensity of the best-fit model (Tab.~\ref{TabParameterFSTauB}). The wavelengths of the observed/modeled NIR and MIR maps are marked at the dominating radiation source.}
\label{SED_bestes_Modell_Anteile} 
 \end{figure}
\subsection{Spectral energy distribution}
\label{AbschnSED12}
The photometry of the best-fit model agrees well with the observational data from the MIR to millimeter wavelengths (Fig.$\,$\ref{SED_bestes_Modell}). The SED shows the characteristic silicate absorption feature at $\sim\unit[10]{\text{\textmu} m}$ which is slightly more pronounced in the modeling than in the IRS dataset. Since the flux in this band is in general very sensitive to changes in disk opacity, a small modification of dust properties, mass, or inclination influences the strength of the band. The simulated fluxes at wavelengths in the MIR beyond the silicate feature are lower than observations, showing deviations up to $40\,\%$. The millimeter data and in particular the spectral index are reproduced quite well. The largest deviations occur at the shortest wavelengths, i.$\,$e. in the optical and NIR. Obviously, the fluxes of the model without extinction screen are too high in this wavelength range. The SED of the model with extinction screen shows a much higher conformity with the observational data. However, the slope in the optical range is too steep and the simulated fluxes at $\lambda=\unit[0.55]{\text{\textmu} m}$ and $\unit[0.7]{\text{\textmu} m}$ are lower than the observational data. In particular, the flux at $\lambda=\unit[0.55]{\text{\textmu} m}$ cannot be explained by the model.

The percentage of scattered stellar radiation and thermal dust \mbox{re}emission is presented in Figure~\ref{SED_bestes_Modell_Anteile} as a function of wavelength. Thermal \mbox{re}emission comprises radiation emitted and possibly scattered by dust grains. Increasing wavelength leads to reduced scattering, thus thermal \mbox{re}emission dominates at wavelengths $\lambda\gtrsim\unit[3]{\text{\textmu} m}$. Therefore, the NIR~maps of the \textit{HST} consist almost entirely of scattered stellar radiation, while the new \textit{NACO}~observation at $\lambda=\unit[3.74]{\text{\textmu} m}$ (Fig.$\,$\ref{map_Image_LBand}) contains $\unit[80]{\%}$ thermal \mbox{re}emission according to our results.

\subsection{Simulated maps and radial profiles}
\label{AbschnittKarten}
\begin{figure} 
 \includegraphics[trim= 2.5cm 7.4cm 6.4cm 4.4cm,clip=true,width=1.0\linewidth]{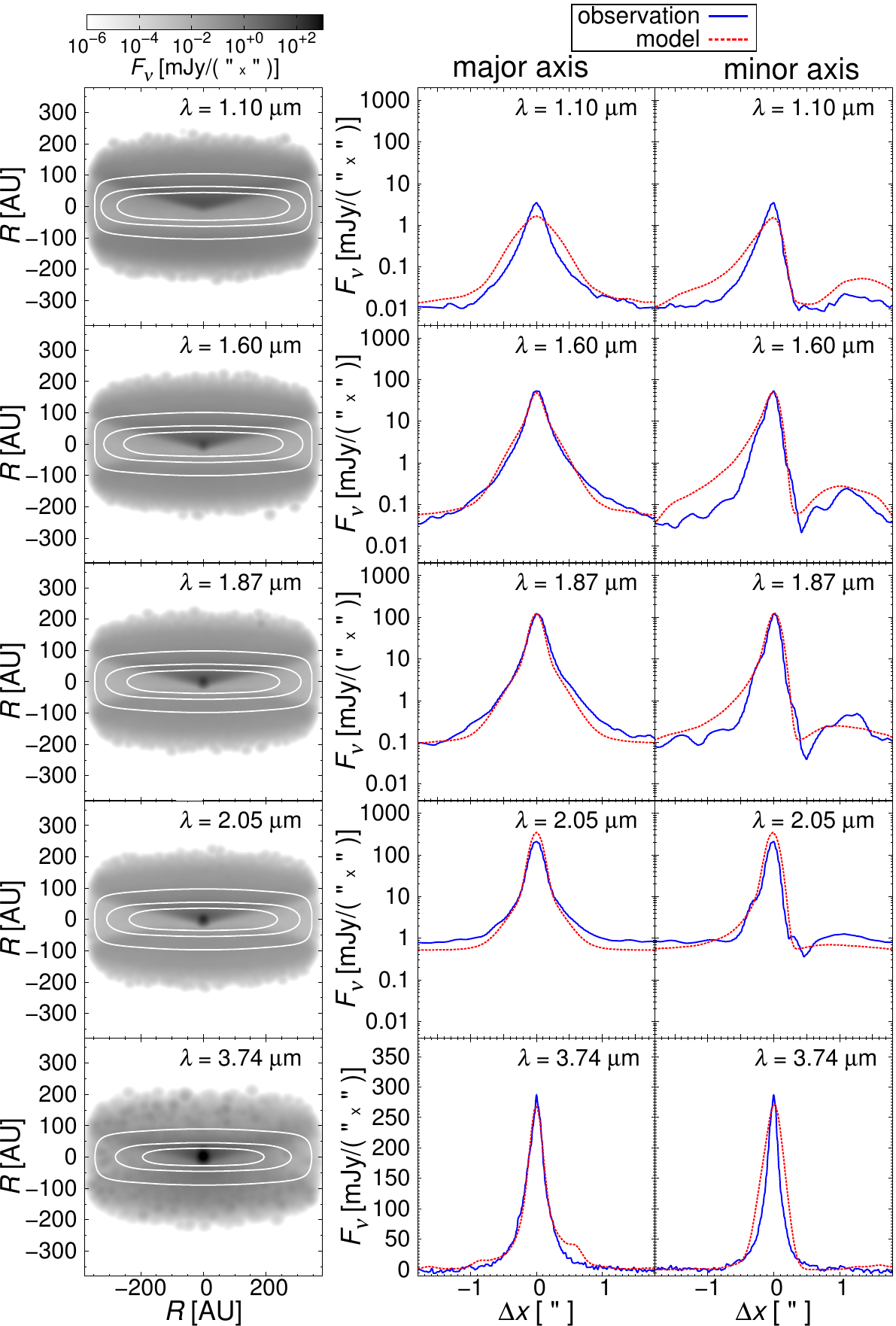}
 	\caption{Left: Synthetic images of the best-fit model of FS~Tau~B at five wavelengths $\lambda$. The solid lines show the surfaces at which the optical depth for an outside observer takes the values $\tau_\lambda=1,\,5$, and 10, respectively. Middle and right: Corresponding radial profiles of FS~Tau~B. The cuts were taken along the major and minor axis. For details see Sect.~\ref{AbschnittKarten}.}
 	\label{FSTAUB_results_maps}
\end{figure}

The synthetic images and their corresponding cuts at $\lambda=\unit[1.10]{\text{\textmu} m}, \unit[1.60]{\text{\textmu} m},\unit[1.87]{\text{\textmu} m}$, $\unit[2.05]{\text{\textmu} m}$, and $\unit[3.74]{\text{\textmu} m}$ are compiled in Figure~\ref{FSTAUB_results_maps}. Apart from asymmetries which cannot be reproduced by the radially symmetric approach of the chosen model, the maps are comparable to the observations.

The profiles along the major axis are well reproduced by the model. Only the simulated profile of the shortest wavelength $\lambda=\unit[1.10]{\text{\textmu} m}$ appears significantly wider than its observational counterpart. The profiles along the minor axis exhibit much larger deviations. The observations show a global maximum and a secondary local maximum, separated by an opaque band in the form of a local minimum. The contrast between the secondary maximum and the minimum of the simulated profiles cannot be reproduced, similar to the modeling of HH~30 (\citealt{Cotera2001}; \citealt{Madlener2012}). Apart from $\lambda=\unit[1.10]{\text{\textmu} m}$, the minimum of the observed profiles is more pronounced than for the simulated profiles. This phenomenon might reveal a systematic deviation of our model from reality. One explanation for this discrepancy lies in a gas and dust stream driven by the central jet which substantially enhances the scattered stellar radiation in this region and hence the contrast (e.$\,$g.,~\citealt{Pety2006}).

An increase of the wavelength results in a decrease of contrast until it vanishes at the longest wavelength. The contrast at the shortest wavelength $\lambda=\unit[1.10]{\text{\textmu} m}$ is low and contrary to the trend. Our model can not explain this observation.

A way to improve the prediction is to modify the underlying dust distribution. In particular a spatial variation of the particle size, caused by dust settling and segregation of large dust particles, has a strong influence on the local chromaticity and thus on scattered and reemitted radiation from the disk (e.$\,$g., \citealt{Pinte2008}; \citealt{Liu2012}; \citealt{Graefe2013}).

 \begin{figure}
 \includegraphics[height=1.0\linewidth, angle=270]{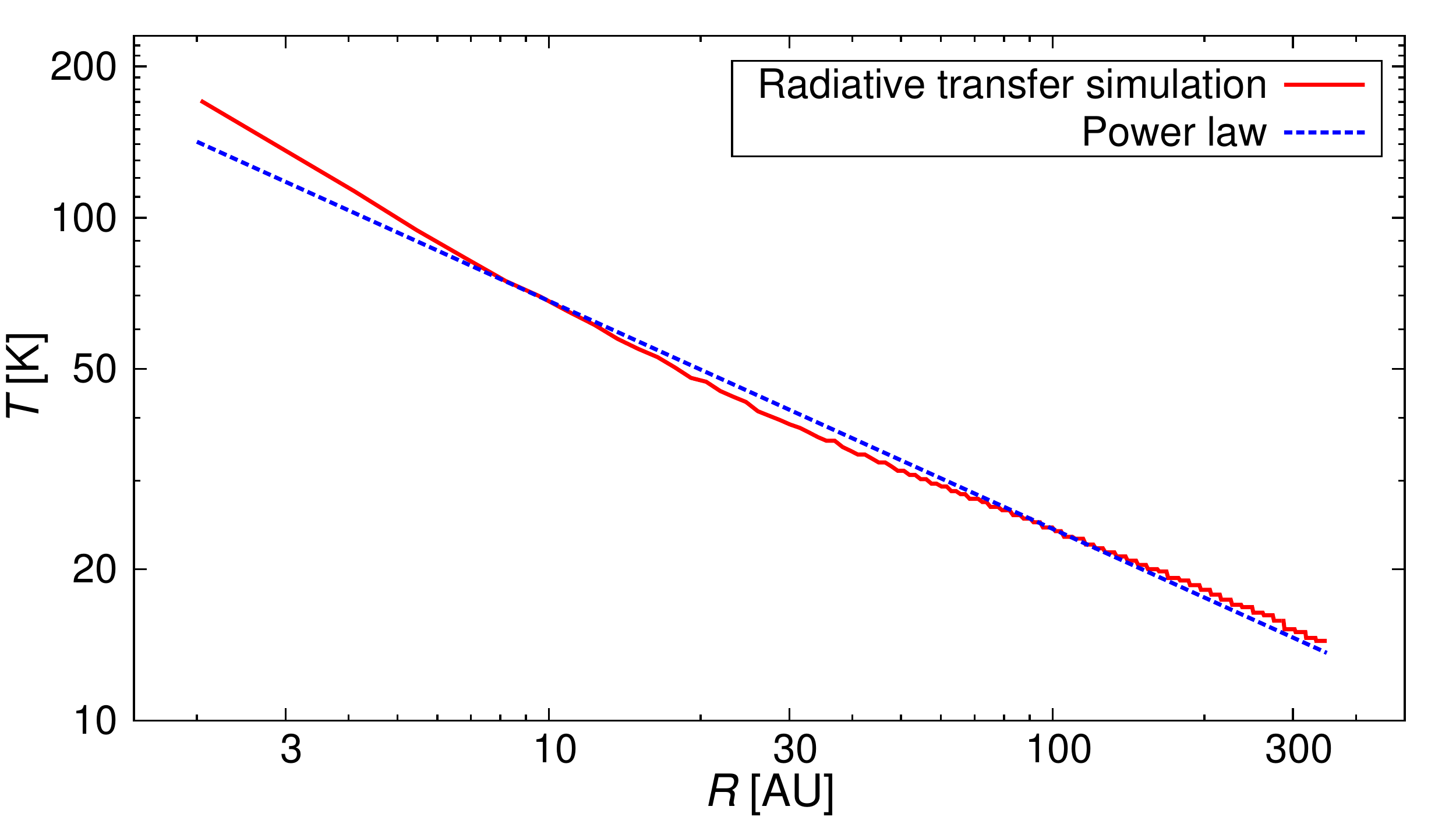}
\caption{Temperature $T$ in disk midplane as a function of the disk radius $R$ for the best-fit model. The red solid line shows the temperature determined in the radiative transfer simulations and the blue dashed line is the fitting by a power law $T(R)=T_{100}\left(\frac{R}{\unit[100]{AU}}\right)^{-q}$ with
$T_{100}=\unit[24.44]{K}\pm\unit[0.07]{K}$ and $q= 0.450\pm0.002 $.}
\label{Midplanetemp_bestes_Modell} 
 \end{figure}

\subsection{Midplane temperature}
\label{Temp2000}
Figure~\ref{Midplanetemp_bestes_Modell} shows the calculated temperature in the disk's midplane for the best-fit model. The temperature varies from $\sim\unit[15]{K}$ in the outer regions up to $\sim\unit[200]{K}$ at the disk's inner edge. A power law 
\begin{align}
T(R)&=T_{100}\left(\frac{R}{R_{100}}\right)^{-q} \label{TempPotenz}
\end{align}
is used to fit the temperature beyond $\unit[4]{AU}$. Using the method of least squares, we derive the values $T_{100}=\unit[24.44]{K}\pm\unit[0.07]{K}$ and $q= 0.450\pm0.002 $. Moreover, the dust mass mean temperature in the disk is $\left\langle T_\text{dust}\right\rangle=\unit[26.7]{K}$. Because the relative deviation of the fit from the simulated distribution is smaller than $\unit[10]{\%}$ for radii $R>\unit[4]{AU}$, the power law represents the simulated temperature quite well. In contrast, at radii below $\unit[4]{AU}$ the temperature shows a steeper slope which can also be observed for other objects (e.$\,$g.,~HH~30, \citealt{Madlener2012}). This inner region is heated in our model by the protostar and the hot inner edge, i.$\,$e. NIR and MIR~radiation provides a contribution to the heating. This is not the case at larger distances, causing the reduction of the temperature slope at several AU. 

 \cite{Yokogawa2001} determined the values $T_{100}=\unit[19.76]{K}$ and $q= 0.61$ which results in a steeper temperature slope. However, a different dust opacity and simpler disk geometry was used in their modeling campaign and only the SED was fitted.

\subsection{Discussion}
In this section we discuss constraints on several disk parameters by the observational data.

The disk inclination is limited by the photometry from $\unit[1-50]{\text{\textmu} m}$ and the shape of the radial brightness profiles of all five maps. The inclined disk does not only shield the stellar scattered radiation by itself, as this scenario yields too high fluxes at optical wavelengths. Only a model of a highly inclined disk combined with an extinction screen yields optical/NIR~fluxes of the same order as the observations. Because of the dependence on both SED and shape of the brightness profiles, the inclination depends only weakly on the value of $A_\text{V}$. Indeed, the quantity $A_\text{V}$ is only constrained by observations in the wavelength range $\unit[0.5-3]{\text{\textmu} m}$.

Because of the high inclination and optical depth, the radial disk structure ($R_\text{in}$, $R_\text{out}$, $\alpha$) is not as well constrained as the vertical density distribution ($\beta, h_{100}$). In particular, the outer radius of several hundreds AU has a large uncertainty interval because it is not traced by the NIR and MIR~maps and only weakly by \mbox{(sub-)}millimeter photometry. Instead, resolved interferometric observations at \mbox{(sub-)}millimeter wavelengths are needed to constrain the outer radius.

The spectral index by itself indicates the presence of large particles. Furthermore, the wavelength range $\unit[1-3]{\text{\textmu} m}$ yields too low fluxes in the scattering regime if only submicrometer particles are considered. Since this influences the profiles in the NIR, the maps constrain the maximum grain radius as well. The scattered fraction for the MIR~map is only $20\,\%$ of the total flux, thus the impact of this map on the maximum grain radius is smaller. The dust mass, however, influences mostly the \mbox{(sub-)}millimeter wavelength domain and the depth of the silicate band.

The geometric parameters describe the spatial distribution of the material which emits at wavelengths beyond the NIR. Therefore, observations in the MIR are important because the contribution of the dust component becomes comparable to or greater than the protostar. The L$_{\text{p}}\,$image shows that the emitting material is located close to the center, thereby putting a limit on the inner radius. The asymmetry of the Lp image is below its spatial resolution. At shorter wavelengths, the inner edge is not directly visible, as can be seen from the asymmetry of the NIR~maps, while the MIR~map is more symmetrical.

The settling of larger grains is not considered in the modeling but might have a strong influence on the scattered and reemitted radiation from the disk (e.$\,$g., \citealt{Pinte2008}; \citealt{Liu2012}) and thus also on disk parameters. In particular, \cite{Graefe2013} found in the case of the Butterfly~star IRAS~04302+2247 a reduction of the scale height $h_{100}$ and steeper slopes of the radial \mbox{(sub-)}millimeter profiles. The decreased incidence of larger grains with several $\text{\textmu}\text{m}$ diameter in upper disk layers leads to a reduction of scattered radiation in the wavelength range of a few $\text{\textmu}\text{m}$ and consequently also on the optical extinction $A_\text{V}$. In general, the short-wavelength range and the slopes of the radial profiles are affected by dust settling, while the influence at millimeter wavelengths is minor.


\section{Conclusion}
In this paper a spatially resolved observation of FS~Tau~B obtained with the instrument \textit{NACO/VLT} in the MIR (L$_\text{p}$~band, $\lambda=\unit[3.74]{\text{\textmu} m}$) was presented. Based on this new image, previously published photometry, and a spatially resolved observation in the NIR taken with \textit{NICMOS/HST}, a parameter study was performed which resulted in new constraints for the disk parameters. The main observables are reproduced satisfactorily by the best-fit model.

The disk extends from an inner radius at $R_\text{in}=\unit[2]{AU}$ to an outer radius of several hundreds AU. The values for the scale height at radius $R_{100}=\unit[100]{AU}$, $h_{100}=\unit[10_{-1}^{+\,2}]{AU}$, and the geometrical parameters $\alpha=2.1_{-0.6}^{+\,0.5}$ and $\beta=1.20_{-0.01}^{+\,0.06}$ are found to be in the typical range for protoplanetary disks. Moreover, the surface density decreases moderately with $p=0.9_{-0.6}^{+\,0.5}$. In summary, parameters describing the vertical density distribution ($\beta$, $h_{100}$) of the disk are better constrained than those influencing the radial disk structure ($R_\text{in}$, $R_\text{out}$, $\alpha$). The temperature in the midplane at $R=\unit[100]{AU}$ has a value of $T_{100}\approx\unit[24]{K}$.

The dust mass is determined to $M_\text{dust}=\unit[2.8\times10^{-4}]{M_\odot}$. Assuming the canonical ratio of gas to dust, $M_\text{gas}/M_\text{dust}=100$ (e.$\,$g.,~\citealt{Hildebrand1983}), we derive a total disk mass of \mbox{$M_\text{disk}=\unit[2.8\times10^{-2}]{M_\odot}$}. 
Evaluation of Toomre's criterion suggests gravitational stability throughout the disk. To reproduce the observational data much larger dust grains ($a_\text{max}=\unit[1]{mm}$) than primordial particles of the ISM are needed. The spectral index $\alpha_\text{mm}\approx2.6$ implies the presence of larger dust particles and therefore grain growth in the disk. The inclination $i=80^\circ$ is constrained by a combination of SED and images, and an extinction screen in the foreground with an optical extinction of $A_\text{V}=12$ gives the best results. The heating source in our best-fit model has a luminosity of $L_\star\approx\unit[9.5]{L_\odot}$. The mass accretion rate is derived to $\unit[3.2\times10^{-7}]{\nicefrac{\text{M}_\odot}{yr}}-\unit[1.2\times10^{-6}]{\nicefrac{\text{M}_\odot}{yr}}$.

The observed SED is well reproduced by the presented best-fit model. While the NIR~maps observed with \mbox{\textit{NICMOS/HST}} consist almost entirely of scattered stellar radiation, the MIR~observation is dominated ($\sim\unit[80]{\%}$) by thermal \mbox{re}emission. The simulated maps show a highly inclined disk. The radial profiles along the major axis are well reproduced, whereas the deviations on the minor axis are larger.

The modeling is based on spatially resolved NIR~and MIR~observations. The decreased optical depth in the MIR reveals slightly deeper embedded regions and potentially larger particles which settled towards the disk midplane and moved closer to the central star. In our study we need the presence of larger particles for the modeling but the resolution of the MIR map is too low to get hints for dust settling or to exclude the occurrence of larger particles in the disk's surface regions.

In general, observations at longer wavelengths probe larger grain sizes and deeper disk regions. Therefore, future studies have to verify the presented model by taking into account observations in the far-infrared and at \mbox{(sub-)}millimeter wavelengths. In addition, the spatial variation of particle size within the disk and thus the spatial dependency of the spectral index can be investigated with observations at these wavelengths. \textit{ALMA}, the largest \mbox{(sub-)}millimeter interferometer (e.$\,$g.,~\citealt{Boley2012}) and other high resolution and sensitive observatories, such as the planned \textit{JWST} (e.$\,$g.,~\citealt{Mather2010}), will enable us to investigate the disk of FS~Tau~B with increased sensitivity on smaller scales, and to obtain a better understanding of disk evolution in general. 

\section*{Acknowledgements}

F.K. acknowledges financial support by the German Research Foundation (\textsl{Deutsche Forschungsgemeinschaft, DFG}) through the project WO 857/7-1. We wish to thank Yao Liu.
\appendix
\setcounter{secnumdepth}{+2}
\setcounter{section}{0} 
\renewcommand\thesection{\Alph{section}}
\section{Photometric data for FS~Tau~B}
\newpage
\begin{table}
\caption{Photometric data points for FS~Tau~B which are considered in the disk modeling (top) or are rejected (bottom).}
\label{Table_SEDFSTauB} 
\begin{tabular}{c c c c l} 
\hline\hline 
 \hspace*{-0.2cm}$\lambda~[\text{\textmu}$m$]$ 	& $F_\nu~[\text{mJy}]$		 & $\sigma~[\text{mJy}]$	 & Instrument & \hspace*{-0.2cm}Ref. \\ \hline 
 \hspace*{-0.2cm}\phantom{0000}0.55\phantom{0}	&	\phantom{$\le$0000}0.0354		&\phantom{000}0.0034 	 & \small{\textit{WFPC2/HST}} & (a) \\
 \hspace*{-0.2cm}\phantom{0000}0.7\phantom{00}	&	\phantom{$\le$0000}0.0456		&\phantom{000}0.0044 	 & \small{\textit{WFPC2/HST}} & (a) \\
 \hspace*{-0.2cm}\phantom{0000}0.9\phantom{00}	&	\phantom{$\le$0000}0.118\phantom{0}	&\phantom{000}0.012\phantom{0} & \small{\textit{WFPC2/HST}} & (a) \\
 \hspace*{-0.2cm}\phantom{0000}1.1\phantom{00}	&	\phantom{$\le$0000}0.77\phantom{00}	&\phantom{000}0.03\phantom{00} & \small{\textit{NICMOS/HST}} & (b) \\
 \hspace*{-0.2cm}\phantom{0000}1.6\phantom{00}	&	\phantom{$\le$0000}5.79\phantom{00}	&\phantom{000}0.11\phantom{00} & \small{\textit{NICMOS/HST}} & (b) \\
 \hspace*{-0.2cm}\phantom{0000}1.87\phantom{0}	&	\phantom{$\le$000}10.9\phantom{000}	&\phantom{000}0.2\phantom{000} & \small{\textit{NICMOS/HST}} & (b) \\
 \hspace*{-0.2cm}\phantom{0000}2.05\phantom{0}	&	\phantom{$\le$000}14.6\phantom{000}	&\phantom{000}0.1\phantom{000} & \small{\textit{NICMOS/HST}} & (b) \\
 \hspace*{-0.2cm}\phantom{0000}2.159		&	\phantom{$\le$000}12.8\phantom{000}	&\phantom{000}1.28\phantom{00} & \small{\textit{QUIRC/MKO}} & (c) \\
 \hspace*{-0.2cm}\phantom{0000}3.6\phantom{00}	&	\phantom{$\le$000}42.116\phantom{0}	&\phantom{000}3.939\phantom{0} & \small{\textit{IRAC/SST}} & (d) \\
 \hspace*{-0.2cm}\phantom{0000}3.74\phantom{0}	&	\phantom{$\le$000}53.6\phantom{000}	&\phantom{000}1.1\phantom{000} & \small{\textit{NACO/VLT}} & (e) \\
 \hspace*{-0.2cm}\phantom{0000}4.5\phantom{00}	&	\phantom{$\le$000}83.587\phantom{0}	&\phantom{000}3.123\phantom{0} & \small{\textit{IRAC/SST}} & (d) \\
 \hspace*{-0.2cm}\phantom{0000}5.8\phantom{00}	&	\phantom{$\le$00}154.499\phantom{0} 	&\phantom{000}9.447\phantom{0} & \small{\textit{IRAC/SST}} & (d) \\
 \hspace*{-0.2cm}\phantom{0000}8.0\phantom{00}	&	\phantom{$\le$00}276.857\phantom{0}	&\phantom{00}16.929\phantom{0} & \small{\textit{IRAC/SST}} & (d) \\
 \hspace*{-0.2cm}\phantom{000}11.56\phantom{0} &	\phantom{$\le$00}429.0\phantom{000}	&\phantom{000}2.491\phantom{0} & \small{\textit{IRS/SST}} & (f) \\
 \hspace*{-0.2cm}\phantom{000}16.356		&	\phantom{$\le$00}946.084\phantom{0}	&\phantom{000}3.359\phantom{0} & \small{\textit{IRS/SST}} 	& (f) \\
 \hspace*{-0.2cm}\phantom{000}24.0\phantom{00}	&	\phantom{$\le$0}1776\phantom{.0000}	&\phantom{0}172\phantom{.0000} & \small{\textit{MIPS/SST}}	& (f) \\
 \hspace*{-0.2cm}\phantom{000}31.597		&	\phantom{$\le$0}2502.883\phantom{0}	&\phantom{000}7.235\phantom{0} & \small{\textit{IRS/SST}} & (f) \\
 \hspace*{-0.2cm}\phantom{000}37.186		&	\phantom{$\le$0}3046.1\phantom{000}	&\phantom{00}26.807\phantom{0} & \small{\textit{IRS/SST}} & (f) \\
 \hspace*{-0.2cm}\phantom{0}1300\phantom{.000}	&	\phantom{$\le$00}141\phantom{.0000}	&\phantom{00}13\phantom{.0000} & \small{\textit{MPIfRBS/SEST}} & (g) \\
 \hspace*{-0.2cm}\phantom{0}2000\phantom{.000} &	\phantom{$\le$000}36\phantom{.0000}	&\phantom{000}8\phantom{.0000} & \small{\textit{NMA}} & (h) \\
 \hspace*{-0.2cm}\phantom{0}2126\phantom{.000} &	\phantom{$\le$000}36.7\phantom{000}	&\phantom{000}2.6\phantom{000} & \small{\textit{NMA}} & (i) \\
 \hspace*{-0.2cm}\phantom{0}2700\phantom{.000}	&	\phantom{$\le$000}22\phantom{.0000}	&\phantom{000}3\phantom{.0000} & \small{\textit{PdBI/IRAM}} & (j) \\\hline
\multicolumn{5}{l}{\hspace*{-0.2cm}Not considered in the disk modeling:}\\
 $2\times10^4\phantom{00}$ 	 		&	\phantom{$\le$0000}0.684\phantom{0}	&\phantom{000}0.123\phantom{0} & \small{\textit{VLA}} & (k) \\
 $6\times10^4\phantom{00}$ 	 		&	\phantom{$\le$0000}0.159\phantom{0}	&\phantom{000}0.108\phantom{0} & \small{\textit{VLA}} & (k) \\\hline
\end{tabular}
\textbf{Note:} (a) \cite{Krist1998}; (b) \cite{Padgett1999}; (c) \cite{Connelley2007}; (d) \cite{Luhman2010}; (e) this work; (f) \cite{Furlan2011}; (g) \cite{Reipurth1993}; (h) \cite{Kitamura2002}; (i) \cite{Yokogawa2001}; (j) \cite{Dutrey1996}; (k)~\cite{Brown1986}.
\end{table}

\bibliographystyle{mnras}
\bibliography{FSTauB} 

\bsp	
\label{lastpage}
\end{document}